\documentclass[aip,jcp,reprint,superscriptaddress,amsmath,amssymb]{revtex4-1}

\usepackage{bm}
\usepackage{amsmath}
\usepackage{graphicx}
\usepackage{dcolumn}
\begin{document}
\title{A fingerprint based metric for measuring similarities of crystalline structures}



\author{Li Zhu}
\affiliation{Department of Physics, Universit\"at Basel, Klingelbergstr. 82, 4056 Basel, Switzerland}
\author{Maximilian Amsler}
\affiliation{Department of Materials Science and Engineering, Northwestern University, Evanston, IL 60208, USA}
\affiliation{Department of Physics, Universit\"at Basel, Klingelbergstr. 82, 4056 Basel, Switzerland}
\author{Tobias Fuhrer}
\affiliation{Department of Physics, Universit\"at Basel, Klingelbergstr. 82, 4056 Basel, Switzerland}
\author{Bastian Schaefer}
\affiliation{Department of Physics, Universit\"at Basel, Klingelbergstr. 82, 4056 Basel, Switzerland}
\author{Somayeh Faraji} 
\affiliation{Institute for Advanced Studies in Basic Sciences, P.O. Box 45195-1159, Zanjan, Iran}
\author{Samare Rostami}
\affiliation{Institute for Advanced Studies in Basic Sciences, P.O. Box 45195-1159, Zanjan, Iran}
\author{S. Alireza Ghasemi}
\affiliation{Institute for Advanced Studies in Basic Sciences, P.O. Box 45195-1159, Zanjan, Iran}
\author{Ali Sadeghi} 
\affiliation{Physics Department, Shahid Beheshti University, G. C., Evin, 19839 Tehran, Iran}
\author{Migle Grauzinyte}
\affiliation{Department of Physics, Universit\"at Basel, Klingelbergstr. 82, 4056 Basel, Switzerland}
\author{Christopher Wolverton}
\affiliation{Department of Materials Science and Engineering, Northwestern University, Evanston, IL 60208, USA}
\author{Stefan Goedecker}
\email{stefan.goedecker@unibas.ch}
\affiliation{Department of Physics, Universit\"at Basel, Klingelbergstr. 82, 4056 Basel, Switzerland}

\begin{abstract}
Measuring similarities/dissimilarities between atomic structures is important for the exploration of 
potential energy landscapes. However, the cell vectors together with the coordinates of the atoms, which are generally used to describe periodic systems, are quantities not suitable as fingerprints to distinguish structures. Based on a characterization of 
the local environment of all atoms in a cell we introduce crystal fingerprints that can be calculated 
easily and allow to define configurational distances between crystalline 
structures that satisfy the mathematical 
properties of a metric. This distance between two configurations is a measure of their
similarity/dissimilarity and it allows in particular to distinguish structures. The new method is an useful tool within various energy 
landscape exploration schemes, such as minima hopping, random search, swarm intelligence algorithms and 
high-throughput screenings.
\end{abstract}

\pacs{61.90.+d, 31.15.A}

\maketitle

\section{INTRODUCTION}
Large data sets of crystalline structures are nowadays available in two major contexts. 
On one hand, databases of materials have been created containing structural information
of both experimental and theoretical compounds from high-throughput calculations, which are the basis for data-mining techniques 
in materials discovery projects~\cite{morgan_high-throughput_2005,saal_materials_2013,curtarolo_aflow:_2012,Jain2013,de_jong_charting_2015,qu_electrolyte_2015}.
On the other hand, ab initio structure 
predictions~\cite{Goedecker:2004kt,Amsler:2010ek,Glass:2006io,Pickard:2011bu,Wang:2010ef,Wang:2012dn,Schon:2001jr,woodley} 
can produce a huge number of new structures that 
 have either not yet been found experimentally or are metastable~\cite{Amsler:2012cq,Amsler:2012co,Amsler:2013hz,Zhu:2012cc,Zhu:2014hy,Zhang:2013be}.
In both cases it is essential to quantify similarities and dissimilarities between structures in the data 
sets, requiring a configurational distance that satisfies the properties of a 
metric. Databases frequently contain duplicates and insufficiently characterized structures which
need to be identified and filtered. In experimental data, the representation of identical 
structures as obtained from different experiments will always slightly differ due to noise in the measurements, such that the 
configurational distance is never exactly zero. Noise is also present in theoretical calculations where a geometry relaxation is for instance 
stopped once a certain, possibly insufficient convergence threshold is reached. 
In ab initio structure prediction schemes it is typically necessary to maintain some structural 
diversity which can be quantified  as a certain minimal configurational distance. 
All these examples clearly show the need for a metric that allows to measure configurational distances and local structures
in a reliable and efficient way.

Crystalline structures are typically given in a dual representation. The first part specifies the cell and the second part 
the atomic positions within the cell.  The former can for instance be given by the three lattice vectors $\mathbf{a}$, $\mathbf{b}$ and $\mathbf{c}$, or by their lengths $a$, $b$ and $c$, and the intermediate angles $\alpha$, $\beta$ and $\gamma$. 
The atomic positions can either be specified by cartesian coordinates or the reduced coordinates with respect to the lattice vectors.
However, such representations are not unique, since any choice of lattice points can serve as cell vectors
of the same crystalline structure. Unique and preferably standardized cell parameters are required for comparison and analysis of different crystals~\cite{lonie_identifying_2012}. Algorithms
to transform unit cells to a reduced form  are frequently used in crystallography, such as the Niggli-reduction~\cite{niggli,Grosse-Kunstleve:sh5006}
which produces cells  with shortest possible vectors ($ | \mathbf{a}+\mathbf{b}+\mathbf{c} | =\textrm{minimal}$). Unfortunately, 
in the presence of noisy lattice vectors, cells can change discontinuously within the Niggli-reduction algorithm. 
Symmetry analysis and the corresponding classification in the 230 crystallographic space groups are another tool to compare crystal structures.
However, the outcome of a symmetry analysis algorithm strongly depends on a tolerance parameter such that the introduction of some noise can change the resulting space group in a discontinuous manner.
Because of the above described problems it is difficult to quantify similarities based on dual representations.

Within the structure prediction community fingerprints that are not based on such a dual representation have been proposed. 
Oganov et al.~\cite{Oganov:2009gr} introduced element resolved radial distribution functions as a crystal fingerprint. For a crystal containing 
one element only a single function is obtained for the entire system. The difference between the radial distribution 
functions of two crystals is then taken as the configurational distance. By definition the radial distribution 
function contains only radial information, but no information about the angular distribution of the atoms. 
Such angular information has been added in the bond characterization matrix (BCM)  fingerprint~\cite{Wang:2012dn,Wang:2015ev}. 
In this fingerprint spherical harmonic and exponential functions are used to set up modified bond-orientational 
order metrics~\cite{Steinhardt:1983hz} of the entire configuration. 
The distance between two configurations can be measured by the Euclidean distance between their BCMs. 
Atomic environment descriptors are also needed in the context of machine learning schemes for force 
fields~\cite{Behler2011,Csanyi,Lilienfeld}, bonding pattern recognition~\cite{ceriotti}, or to compare vacancy, interstitial and intercalation sites~\cite{yang_proposed_2014}. 
These descriptors could also be used to measure similarities between structures. Even though 
they have never been used in this context we will present a comparison with such a descriptor.

When humans decide by visual inspection  whether two structures are similar they proceed typically in a different way.
They try to find matching atoms which have the same structural environment. If all the atoms in one structure can 
be matched with the atoms of the other structure, the two structures are considered to be identical. 
Such a matching approach based on the Hungarian algorithm~\cite{Kuhn:1955hh} has already turned out to be useful for the distinction of 
clusters~\cite{Calaminici, Sadeghi:2013bh}.

In this paper we will present a fingerprint for crystalline structures which is based on such a matching approach. The environment of each atom is 
described by an atomic fingerprint which is calculated in real space for an infinite crystal and represents some 
kind of environmental scattering properties observed from the central atom. Therefore, all the ambiguities of a dual 
representation do not enter into the fingerprint, allowing an efficient and precise comparison of structures.

\section{FINGERPRINT DEFINITION}
Recently we have proposed an configurational fingerprint for clusters~\cite{Sadeghi:2013bh}. In this approach an overlap matrix is calculated for
an atom centered Gaussian basis set. The vector formed by the eigenvalues of this matrix forms a global fingerprint that characterizes 
the entire structure. 
The Euclidian norm of the difference vector between two structures is the configurational distance between  them and satisfies the 
properties of a metric. 

Since there is no unique representation of a crystal by a group of atoms (e.g. the atoms in some unit cell) we will use atomic fingerprints 
instead of global fingerprints in the crystalline case. 
However, this atomic fingerprint is closely related to our global fingerprint for non-periodic systems. 
For each atom $k$ in a crystal located at ${\bf R}_k$ we obtain a cluster of atoms by considering  only those contained in a sphere 
centered at ${\bf R}_k$. 
For this cluster we calculate the overlap matrix elements $S^k_{i,j}$ 
as described in reference~\onlinecite{Sadeghi:2013bh} for a non-periodic system, 
i.e we put on each atom one or several 
Gaussian type orbitals and calculate the resulting overlap integral. The orbitals are indexed by the letters $i$ and $j$ 
and the index $w(i)$ gives the index of the atom on which the Gaussian $G_i({\bf r})$ is centered, i.e. 
\begin{equation}
	S^k_{i,j} = \int d{\bf r} \:\:  G_i({\bf r}-{\bf R}_{w(i)}) \:  G_j({\bf r}-{\bf R}_{w(j)})
\end{equation}
In this first step, the amplitudes of the Gaussians $c_{\textrm{norm}}$ are chosen such that the Gaussians are normalized to one.
To avoid that the eigenvalues have discontinuities when an atom enters into or leaves the sphere we construct in a second step another matrix $T^k$ such that 
\begin{equation} \label{cutoff}
 T^k_{i,j} =  f_c(|{\bf R}_{w(i)}-{\bf R}_k|)   S^k_{i,j} f_c(|{\bf R}_{w(j)}-{\bf R}_k|)  
\end{equation}
The cutoff function $f_c$ smoothly goes to zero on the surface of the sphere with radius 
$\sqrt{2 n} \sigma_c$
\begin{equation}
\label{fcut}
f_c(r) = \left(1- \frac{r^2}{2 n \sigma_c^2}\right)^n
\end{equation}
In the limit where $n$ tends to infinity the cutoff function converges to a Gaussian of 
width $\sigma_c$. The characteristic length scale $\sigma_c$ is typically chosen to be the sum of the 
two largest covalent radii in the system.

The value $n$ determines how many derivatives of the cutoff function are continuous on the surface of the sphere, and $n=3$ was used in the following.
One can consider the modified matrix $T^k$ to be the overlap matrix of the cluster where the amplitude of the Gaussian at atom 
$i$ is determined by $c_{\textrm{norm}} \: f_c(|{\bf R}_i-|{\bf R}_k|)$. In this way atoms close to the surface of the sphere 
give rise to very small eigenvalues of $T^k$ and are thus weighted less than the atoms closer to the center. 
The eigenvalues of this matrix $T^k$ are sorted in descending order and form the atomic fingerprint vector ${\bf V}_k$.
Since we can not 
predict exactly how many atoms will be in the sphere we estimate a maximum length for the atomic fingerprint vector. If the number of 
atoms is too small to generate enough eigenvalues to fill up the entire vector, the entries at the end of the fingerprint vector are filled up with zeros.
This also guarantees that the fingerprint is a continuous function with respect to the motion of the atoms
when atoms might enter or leave the sphere. If an atom enters into the 
sphere some zeros towards the end of the fingerprint vector are transformed in a continuous way into some very small entries which only contribute little to the overall fingerprint. 
The Euclidean norm $| {\bf V}_k - {\bf  V}_l |$ measures the dissimilarity between the atomic environments of atoms $k$ and $l$. 

The atomic fingerprints ${\bf V}^p_k$ and ${\bf V}^q_k$ of all the $N_{at}$ atoms in two crystalline configurations $p$ and $q$ can 
now be used to define a configurational distance $d(p,q)$  between the two crystals:
\begin{equation} \label{def}
	d(p,q)=\underset{P}{\textrm{min}}\left( \sum_k^{N_{at}} | {\bf V}^p_k - {\bf V}^q_{P(k)} |^2 \right)^{1/2},
\end{equation}
where $P$ is a permutation function which matches a certain atom $k$ in crystal $p$ with atom $P(k)$ in crystal $q$. 
The optimal permutation function which minimizes  $d(p,q)$ can be found with the Hungarian algorithm~\cite{Kuhn:1955hh}  in polynomial time. 
If the two crystals $p$ and $q$  are identical the Hungarian algorithm will in this way assign corresponding atoms to each other.
The Hungarian algorithm needs as its input only the cost matrix $C$ given by 
$$ C_{k,l} =  | {\bf V}^p_k - {\bf V}^q_l |^2 $$

In the following it will be shown that $d(p,q)$ satisfies the properties of a metric,
namely
 \begin{itemize}
 \item  positiveness: $d(p,q) \geq 0$ 
 \item  symmetry: $d(p,q) = d(q,p)$
 \item  coincidence axiom: $  d(p,q)=0$  if and only if  $p=q$  
 \item  triangle inequality: $ d(p,r) + d(r,q)  \geq d(p,q) $.
 \end{itemize}

 From the definition (Eq.~\ref{def}) it is obvious that the positiveness and symmetry conditions are fulfilled.
 The coincidence theorem is satisfied if the individual atomic fingerprints are unique, i.e if there are not two 
 different atomic environments that give rise to identical atomic fingerprints. In our work on fingerprints for clusters we 
 have shown that the fingerprints can be considered to be unique if they have a length larger or equal to 3 per atom.
 The triangle inequality can be established in this way:
\begin{align*} 
	    d(p,r) + d(r,q)
		    &=\left( \sum_k^{N_{at}} | {\bf V}^r_k - {\bf V}^p_{P(k)} |^2 \right)^{1/2}  \\
		    &+\left( \sum_k^{N_{at}} | {\bf V}^r_k - {\bf V}^q_{P'(k)} |^2 \right)^{1/2} \\
		    &\geq \left( \sum_k^{N_{at}} | {\bf V}^p_{P(k)} -  {\bf V}^q_{P'(k)} |^2 \right)^{1/2} \\
		    &\geq \left( \sum_k^{N_{at}} | {\bf V}^p_k - {\bf V}^q_{Q(k)} |^2 \right)^{1/2} \\
		    &= d(p,q)
	\end{align*}
where $P$, $P'$ and $Q$ are assumed to be the permutations that minimize respectively 
the Euclidean vector norms associated to $d(p,r)$, $d(r,q)$  and $d(p,q)$.

\section{Contracted fingerprints} 
Since the ${\bf R}_k$-centered spheres contain typically about 50 atoms, an atomic fingerprint has at least length 50 if only $s$-type 
Gaussian orbitals or length 200 if both $s$ and $p$ orbitals are used. Since a configuration is characterized by 
the ensemble of all the atomic fingerprints of all the atoms in the cell, the amount of data needed to 
characterize a structure is quite large even though it is certainly manageable for crystals with a small number of atoms per unit cell. 
Storage requirements might however become too high in certain cases such as large molecular crystals. 
We will, therefore, introduce contraction schemes that allow to  considerably reduce the amount of data necessary to 
characterize a crystalline structure. Two such schemes will briefly be discussed below.

\subsection{Contractions by properties}

Let us introduce a function $\tau(i)$ that designates a certain property of the Gaussian orbital $i$ and encodes it in form of a 
contiguous integer index.
In case of a multicomponent crystal it can indicate on which kind of chemical element the Gaussians are centered  and whether the orbital is of $s$ or $p$ type. 
The principal vector is thus chopped into pieces whose elements all carry the same value $\tau(i)$.
In the following presentation of numerical results we have always 
considered the central atom to be special, independent of its true chemical type. Having $m$ atomic species in the unit cell and using atomic Gaussian orbitals with a maximum angular momentum $l_{\mathrm{max}}$, $\tau(i)$ runs from 1 to $(m+1)(l_{\mathrm{max}}+1)$. Now we can construct a contracted matrix $t^k$ 
$$ t^k_{\nu,\mu} = \sum_{i,j}   \delta_{\nu,\tau(i)} u^k_i  T^k_{i,j} u^k_j   \delta_{\mu,\tau(j)}  $$
together with its metric tensor $s^k$ 
$$ s^k_{\nu,\mu} = \sum_{i}   \delta_{\nu,\tau(i)} u^k_i  u^k_j   \delta_{\mu,\tau(j)}  $$
 where ${\bf u}^k$ is the principal vector of the matrix $T^k$ of Eq.~\ref{cutoff}. 
 The eigenvalues $\lambda$ of the generalized eigenvalue problem
 $$ t^k {\bf v} = \lambda s^k {\bf v} $$
 form again an atomic fingerprint of length $(m+1)(l_{\mathrm{max}}+1)$ which is much shorter than the non-contracted fingerprint ${\bf V}_k$.

\subsection{Contractions to form molecular orbitals for molecular crystals}

The fingerprints described so far  can in principle also be used for molecular crystals. 
However, the amount of data needed to characterize such crystals can be quite large if the 
molecules forming the crystal contain many atoms. By creating molecular orbitals 
in analogy with standard methods in electronic structure calculations the required amount of data can be considerably reduced. 
The eigenvalues arising from the overlap matrix in this molecular basis set will then 
form a fingerprint for the molecular crystal. The molecular orbitals can be obtained 
in the following way: for each molecule $k$ in our unit cell we cut out a cluster of molecules
within a sphere of a certain radius. For each molecule $\alpha$ in this sphere we set up the 
overlap matrix by putting Gaussian type orbitals on all its constituent atoms. 
Then we calculate for this matrix the eigenvalues and eigenvectors. 
The principal vectors ${\bf W}^{\alpha,\mu}$ belonging to several of the largest eigenvalues $\lambda^{\alpha}_{\mu}$ 
are subsequently used for the contraction:
\begin{equation}
{\cal S}^{k}_{\alpha,\mu;\beta,\nu} = \sum_{i,j}   W^{\alpha,\mu}_i S^{k}_{i,j} W^{\beta,\nu}_j 
\end{equation}
No metric tensor is required since the set of vectors used for the contraction is orthogonal.
The molecular orbitals have characteristic patterns, such that the orbital corresponding to the first principal 
vector has no nodes, while the orbitals of the following principal vectors have increasing number of nodes. 
They are therefore similar to the atomic orbitals of $s$, $p$ and higher angular momentum character, 
which were used for the fingerprints in the ordinary crystals. 
In Fig.~\ref{orbitals} these orbitals are shown for the case of the paracetamol molecule.

By multiplying ${\cal S}$ with some cutoff function as in Eq.~\ref{cutoff} we can then obtain  
molecule  centered 
overlap matrices in this molecular basis which is free of discontinuities with respect to the motion 
of the atoms. In the molecular case the value of the cutoff function depends on 
some short range pseudo-interaction between the central and the surrounding molecules. 
This interaction $U_{k,\alpha}$ between the central molecule $k$ and another molecule $\alpha$ is given by 
\begin{equation}
	U_{k,\alpha} = \sum_{i,j}\left ( 1-\frac{d_{i,j}^2}{2n\left ( \sigma^{van}_{i,j} \right )^2} \right )^n
\end{equation}
where the sum over $i$ runs over all the atoms in the central molecule $k$ and the sum over $j$ over all the 
atoms in the surrounding molecule $\alpha$. $d_{i,j}$ is the distance between the atoms $i$ and $j$ and 
$\sigma^{van}_{i,j}$ is the sum of the van der Waals radii of the two atoms. 
The interaction is taken to vanish beyond its first zero. 
Because of the short range of the interaction, molecules sharing a large surface will be coupled strongly.
The analytical form of the cutoff function is identical to the one for the atomic case (Eq.~\ref{fcut}). 
However, since a cartesian distance between molecules is ill defined, the argument in Eq.~\ref{fcut} is modified.
The scaled distance $r \sigma_c $ between the atoms  is replaced by the normalized interaction between the molecules
$$
\frac{U_{k,\alpha}}{U_{k,k}} 
$$
The eigenvalues of this final overlap matrix form now a 
fingerprint describing the environment of this molecule $k$ with respect to the other molecules. 
To compare two structures this procedure is done for all molecules contained in the corresponding unit cell. 
A configurational distance is calculated then as in Eq.~\ref{def} by using the Hungarian algorithm~\cite{Kuhn:1955hh}.

\section{Application of fingerprint distances to experimental structures}
Structural data found in various material databases is frequently obtained from measurements at different temperatures which results 
in thermal expansion. Similarly, measurements at different pressures or low quality x-ray diffraction patterns can lead to slight
cell distortions. Obviously our fingerprint distances among such expanded or distorted but otherwise identical 
structures are different from zero. For these reasons we have introduced a scheme where the six degrees of freedom associated to the 
 cell are optimized while keeping the reduced atomic coordinates fixed such as to obtain the smallest possible distance to a reference configuration. 
The gradient of our fingerprint distance with respect to the lattice vectors can be calculated 
analytically using the Hellmann Feynman theorem. 
An application of the lattice optimization scheme was applied to a subset of ZrO$_2$ structures taken from the Open Quantum Materials Database (OQMD)~\cite{saal_materials_2013}, as will be discussed in further detail later in the following section. 

\section{Numerical tests}

\begin{figure*}[h] 
\begin{center}
\includegraphics[width=\textwidth]{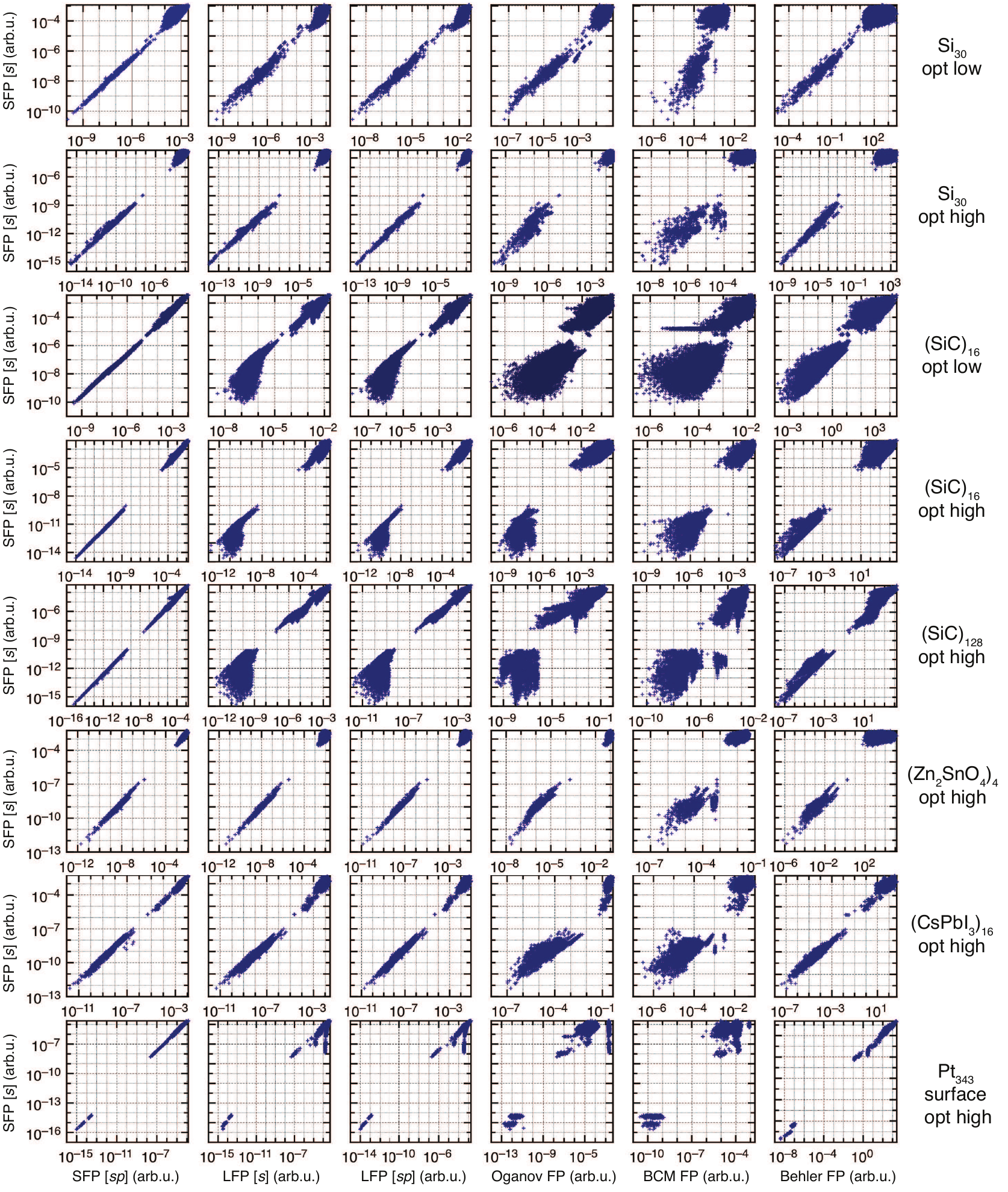}%
\end{center}
\caption{\label{big} Correlation between different fingerprints for all the 8 test sets obtained during structure 
prediction runs. SFP[$x$] and LFP[$x$] indicate short and long fingerprints with $x$ orbitals, respectively. ``opt high'' and ``opt low'' indicate the quality of the geometry relaxation.}
\end{figure*}

Fig.~\ref{big} shows all possible pairwise configurational distances obtained with several fingerprints for various data sets.
Different fingerprints are plotted along the x and y axis. LFP stands for the uncontracted long fingerprint 
and in square parenthesis it is indicated whether only $s$ or both $s$ and $p$ orbitals were used to set up the overlap matrix,
SFP[$s$] stands for the short contracted fingerprint with $s$ orbitals only where the properties used for the contraction 
are central atom and the element type of the neighboring atoms in the sphere.
For materials that have only one type of element (Si in our case) the atomic fingerprint has 
only length two and the coincidence theorem is not satisfied. Even though there are hyperplanes in the configurational space 
where different configurations have identical fingerprints, it is very unlikely that different local 
minima lie on such hyperplanes and the fingerprint can therefore nevertheless well distinguish between 
identical and distinct structures. If both $s$ and $p$ orbitals are used (SFP[$sp$]) the atomic fingerprint has at least length 
4 and no problem with the coincidence theorem arise. In addition we also show the configurational distances arising from the 
Oganov~\cite{Oganov:2009gr} and BCM~\cite{Wang:2012dn,Wang:2015ev} fingerprints as well as from a fingerprint based on the amplitudes 
of symmetry functions~\cite{Behler2011}. 
All our data sets contain both the global minimum (geometric ground state) as well as local minima 
(metastable) structures, obtained from minima hopping runs~\cite{Amsler:2010ek}.  
Energies and forces were calculated with the DFTB+~\cite{dftb} method for SiC and the molecular crystals, and the Lenosky tight-binding scheme was used for Si~\cite{lenosky_highly_1997}. 
For the CsPbI$_3$ perovskite and the transparent conductive oxide Zn$_2$SnO$_4$ plane wave density functional theory (DFT) calculations were used as implemented in the quantum Espresso code~\cite{quantumespresso,ceriotti_i-pi:_2014}. 

The first test set consists of clathrate like structures  of low density silicon allotropes~\cite{amsler2015}. 
Low density silicon gives rise to a larger number of low energy crystalline structures than 
silicon at densities of diamond silicon and thus poses an ideal benchmark system. 
In the first line of the figure we show the results of a relatively sloppy local geometry optimization, 
where the relaxation is stopped once the forces are smaller than 5.e-2~eV/\AA. 
Gaps separating identical from distinct structures are hardly visible for all fingerprints.
Once a very accurate geometry optimization with a force 
threshold of 5.e-3~eV/\AA  $ $ is performed, gaps become visible for all the fingerprints. 

The second data set is silicon carbide, a material well known for its large number of polytypes. 
Our fingerprint gives rise to a small gap whereas the configurational distances based on all other 
fingerprints do not show any gap at all. 
The opening of a gap can again be observed once the geometry optimization is done with high accuracy. 
For this case all fingerprints result in a gap, but like for all test sets it is the least 
pronounced for the BCM fingerprint. 
Both the Oganov and BCM fingerprints are global ones  such that information is lost 
in the averaging process of these fingerprints as the system gets larger. Therefore, it is not surprising that 
the gap again disappears even for the high quality geometry optimization once one goes to large  cells. 

The next two test sets consist of an oxide material and a perovskite with their characteristic building blocks of 
octahedra and tetrahedra which can be arranged in a very large number of different ways.
All our fingerprints give rise to clear gaps separating identical from distinct structures.
The Oganov fingerprint also gives rise to clear gaps whereas the BCM fingerprint only weakly 
indicates some gap. The Behler fingerprint gives a well pronounced gap for Zn$_2$SnO$_4$ but only 
a blurred gap for CsPbI$_3$.

The last theoretical test system is a platinum surface. In this case the energies were calculated with the Morse potential~\cite{morse,Bassett:1978dy}. The geometry optimization were done with high accuracy and therefore a big gap is visible in all cases.

\begin{figure*}[h]
\begin{center}
\includegraphics[width=\textwidth]{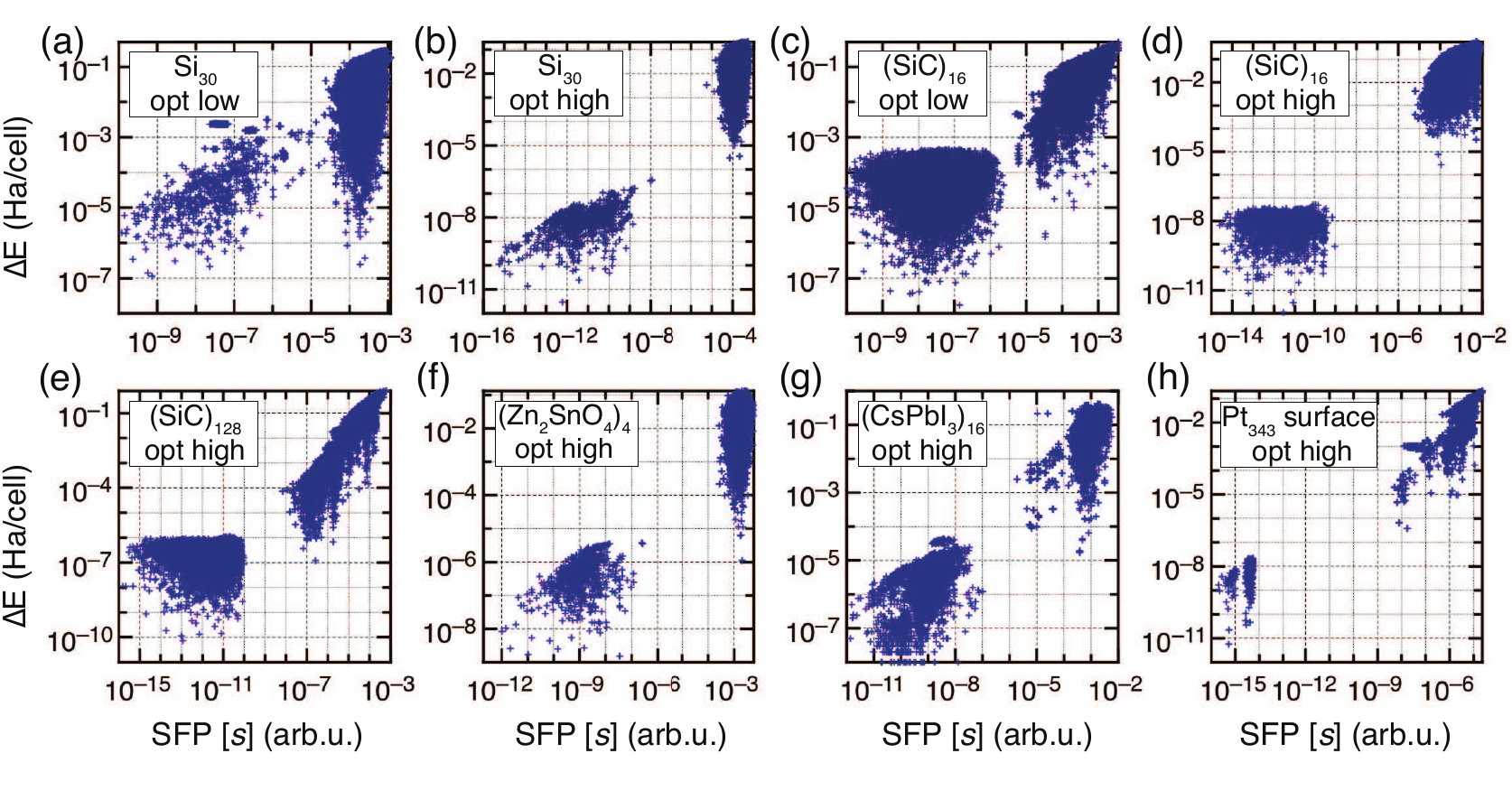}%
\end{center}
\caption{\label{energy} Correlation between the energy difference and the fingerprint distance for all the 8 theoretical test 
setting of Fig~\ref{big}.}
\end{figure*}

Fig.~\ref{energy} shows the correlation between the energy difference and the fingerprint distance for all the 
test cases of Fig.~\ref{big}. Except for the very large 256 atoms system there exists always a clear energy gap 
if the geometry optimization was done with high accuracy. Even though there is of course the possibility 
of nearly degenerate structures, this seems to happen rarely in practice and energy is thus a rather good 
and simple descriptor for small unit cells.

\begin{figure}[h]
\begin{center}
\includegraphics[width=\columnwidth]{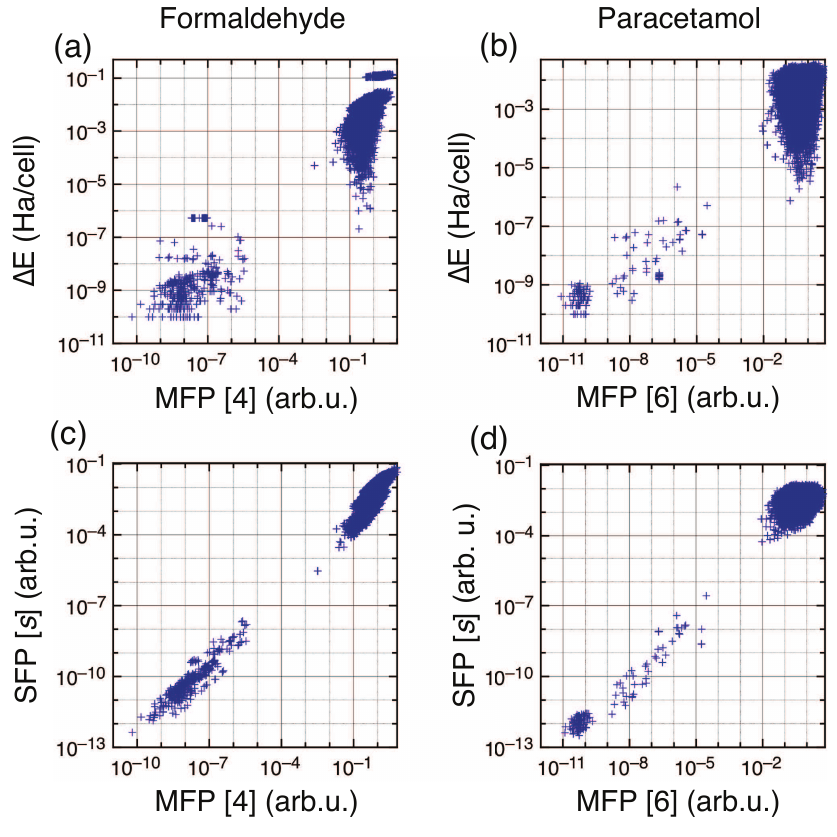}
\end{center}
\caption{\label{molecular} 
Top panels: Correlation between the energy difference and the molecular fingerprint distance (MFP) 
for formaldehyde (a) and paracetamol (b). Bottom panels: Correlation between molecular fingerprint 
distance and standard fingerprint distance (short contracted fingerprint with s orbitals 
only, SFP[s]) for formaldehyde (c) and paracetamol (d). }
\end{figure}

\begin{figure}[h]
\begin{center}
\includegraphics[width=\columnwidth]{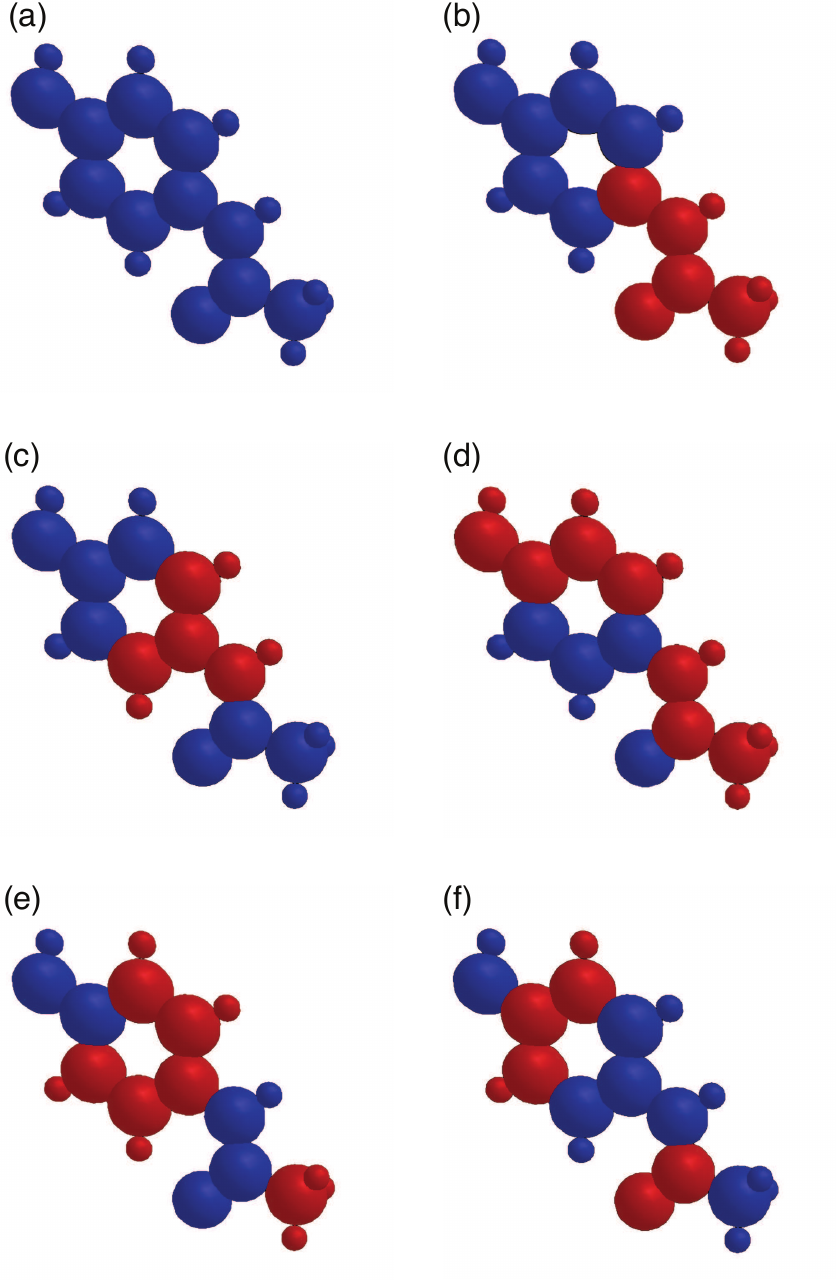}
\end{center}
\caption{\label{orbitals} 
The nodal character of the first six principal vectors for the paracetamol molecule.
The atoms are colored according to the sign 
of the elements of the first six principal vectors ${\bf W}^{\alpha,\mu}$. 
A systematic colour pattern can be observed. The first principal eigenvector never changes sign 
and has therefore no nodes (a). Higher principal vectors exhibit more and more nodes (b-f). }
\end{figure}

To test our molecular fingerprint, two test systems were employed, namely crystalline formaldhyde and paracetamol. 
The formaldehyde system comprised 240 structures with 8 molecules per cell and the paracetamol system 300 structures with 4 molecules per cell. 
The two top panels of Fig.~\ref{molecular} show the molecular fingerprint distance versus the energy difference of different structures 
of paracetamol and formaldehyde, respectively. 
The two bottom panels show the correlation of the standard fingerprint against the molecular fingerprint for both systems.
The existence of a gap in the pairwise distance distributions clearly indicates that identical and distinct structures can 
be identified by both fingerprints. However, the molecular fingerprint vector is considerably shorter because only six principal vectors 
were used (shown in Fig.~\ref{orbitals}). Since six is the number of degrees of freedom of a rigid rotator it is expected that 
this fingerprint is long enough to satisfy the coincidence theorem.

\begin{figure}[h]
\begin{center}
\includegraphics[bb=0 -1 492 346,clip,width=\columnwidth]{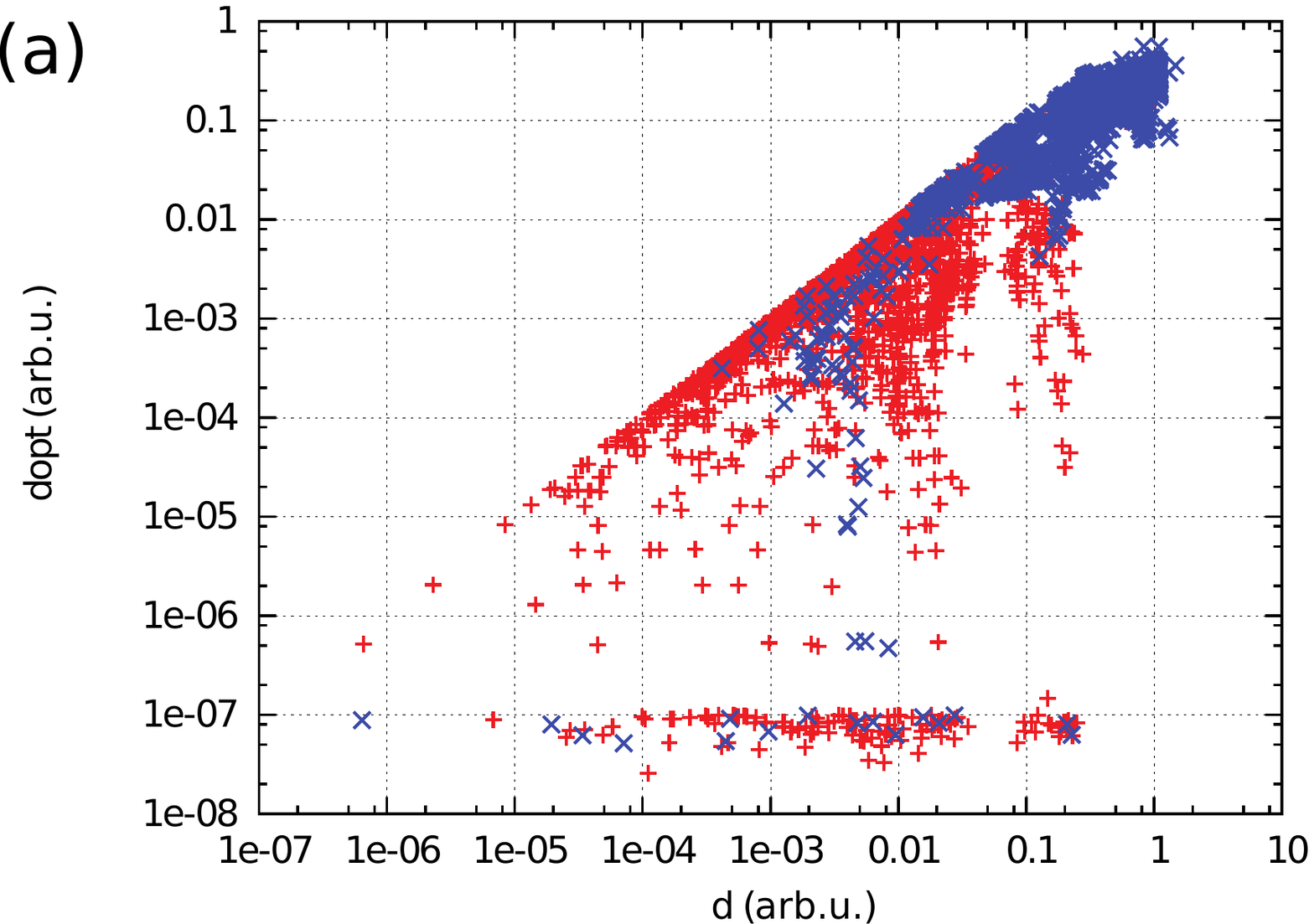}
\includegraphics[bb= 0 -1 484 346,clip,width=\columnwidth]{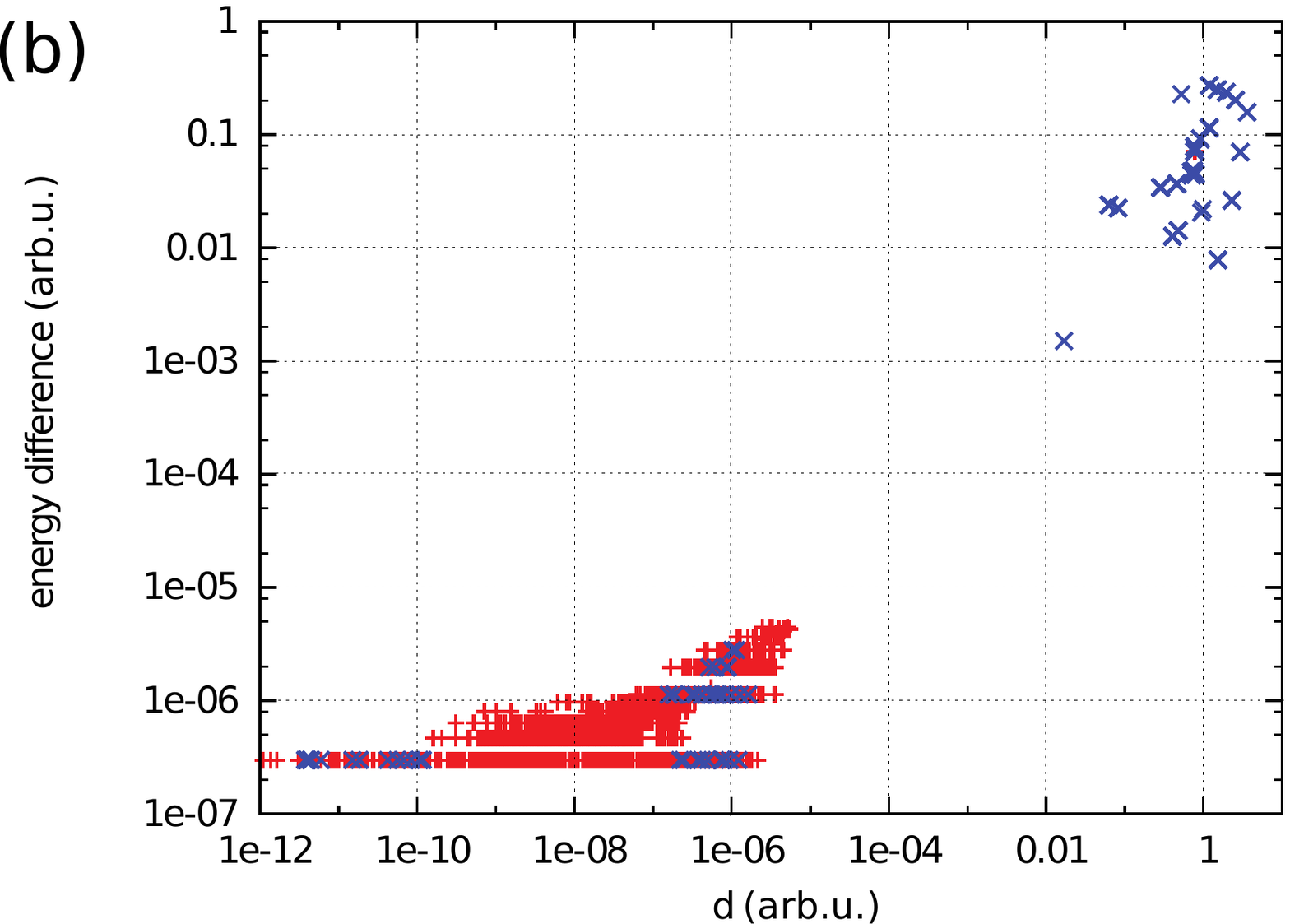}
\end{center}
\caption{\label{ZrO2} 
Panel (a) shows along the x-axis the ordinary fingerprint distances and along the y-axis the lattice optimized
fingerprint distances for the ZrO$_2$ structures retrieved from the OQMD. 
Distances between two structures that were
identified as identical by the structural comparison algorithm implemented in  MINT  are shown in red and structures that were identified as distinct are
shown in blue. Panel (b) shows the correlation between the DFT energy differences among  all 
relaxed structures and the ordinary fingerprint distances.  
}
\end{figure}

Next we applied our fingerprint to ZrO$_2$ structures contained in the OQMD. 
115 different entries were available at this composition. 
The structures were either based on  experimental data retrieved from
the Inorganic Crystal Structure Database (ICSD) or on binary structural prototypes. When the OQMD was initially created, duplicate entries
were identified with the structure comparison algorithm as implemented in the Materials Interface (MINT) software package~\cite{MINT}
which employs a 6-level test that includes cell reduction as well as an analysis of the lattice symmetry.
Structures classified as identical to an existing entry in OQMD were mapped to that entry without performing
a structural relaxation. Therefore, the structural data set contains both 
DFT optimized and experimental structures, resulting in noise on the atomic and cell coordinates arising from the numerical calculations as well as from the different experiments and thermal effects.
In Fig.~\ref{ZrO2}a we show the ordinary and the lattice vector optimized fingerprint distances for all 115 structures 
from the database. We can see that the fingerprint distance can be reduced down to about 1.e-7 for many structures.
For some of them the initial fingerprint distances were as large as 0.1. This allows to detect some identical structures 
whose initial large fingerprint distance was only due to thermal expansion. However, even with lattice vector optimization 
it was not possible to decide for the whole data set in an inambiguous way which structures are identical and which were not.
Therefore, local geometry optimizations were performed at the DFT level for all structures using the VASP code~\cite{vasp,PAW,PAW2}. A plane wave cutoff energy
of 520~eV was used together with a dense $k$-point mesh. Both the atomic and cell variables were relaxed until the maximal force component
was less than 2.e-3~eV/\AA $ $ and the stress below 1.e-2~GPa. Panel (b) of 
Fig.~\ref{ZrO2} shows the DFT energy differences of the relaxed structures against the fingerprint distances, showing a clear gap that allows to distinguish between identical and different structures. Applying the lattice vector optimization scheme on these relaxed structures was not able to further lower the fingerprint distances of identical structures.
The coloring in Fig.~\ref{ZrO2} indicates how the two structures belonging to a fingerprint distance were classified by MINT. Assuming that there are no different structures with degenerate DFT energies, one can conclude that MINT was not able to extract from the non-relaxed data set
the information whether structures are identical or not and has erroneously assigned numerous
identical structures as distinct, and vice versa to a lesser extent.

Since both Oganov and BCM  methods are  global fingerprints that discard crucial information,
they can fail to describe structural differences, a problem that becomes especially apparent 
when considering defect structures in complex materials. As an example, a $2\times2\times2$  supercell was constructed of the cubic perovskite structure
of LaAlO$_3$~\cite{LaAlO3}. Half of the Al atoms on the B-sites were replaced by  
Mn. Then, single oxygen vacancies were introduced on symmetrically inequivalent X-sites. Obviously, the structural symmetry was reduced
from the initial space group  $Pm\bar{3}m$ of LaAlO$_3$ to the orthorhombic space group $Amm2$ of the supercell La(Al,Mn)O$_3$, and the 
oxygen vacancies resulted in structures with $Cm$ and $Pm$ symmetry. Both  MINT and our fingerprint confirm that the structures
are clearly different, whereas the Oganov and BCM fingerprint erronously classify both structures as identical.

\section{Conclusions}
Atomic fingerprints that describe the scattering properties as obtained from an overlap matrix 
are well suited to  characterize atomic environments. 
An ensemble of atomic fingerprints forms a global fingerprint that allows to identify 
crystalline structures and to define 
configurational distances satisfying the properties of a metric. The widely used Oganov and BCM fingerprints 
do not have these properties and do also in practice not allow a reliable way to distinguish identical from distinct structures. 
Symmetry function based fingerprints are of similar quality as our scattering fingerprints.
However, they are much more costly to calculate. Both fingerprints have a cubic scaling with respect to the number of atoms 
within the cutoff range, but our prefactor of the matrix diagonalization is much smaller then the prefactor for 
the 3-body terms required for the calculation of the symmetry functions. 
In contrast to `true'--`false' schemes such as employed in MINT which rely on a threshold and affirm that two structures are 
either identical or distinct, our fingerprint gives a distance between configurations. The appearance of a gap in the distance distribution indicates that a reliable 
assignment of identical and distinct structure can be performed. In addition, strong reductions in the 
fingerprint distances upon lattice vector optimization can detect and eliminate thermal noise on the data set, rendering our fingerprint
ideal to scan for duplicates in large structural databases. 
Our scheme can easily be extended to molecular crystals by introducing quantities that are analogous to 
molecular orbitals. Furthermore, the new fingerprint can be used to accurately explore local environments to create atomic and structural attributes for machine learning techniques.
In summary, we have demonstrated that this approach allows to characterize 
crystalline structures by rather short fingerprint vectors 
and to decide more reliably whether structures are identical or not than previously proposed methods.

\begin{acknowledgments}
We thank Vinay Hegde and Antoine Emery for valuable expert discussions.
This work was done within the NCCR MARVEL project.
MA gratefully acknowledges support from the Novartis
Universit\"at Basel Excellence Scholarship for Life Sciences 
and the Swiss National Science Foundation.
Computer resources were provided at the CSCS under project s499 and the National Energy Research Scientific Computing Center, which is supported by the Office of Science of the U.S. Department of Energy under Contract No. DE-AC02-05CH11231.
 
\end{acknowledgments}

%

\end{document}